\documentclass[aps,twocolumn,prb,preprintnumbers,amsmath,amssymb]{revtex4-1}
\usepackage{graphicx, bm}
\usepackage{dcolumn}
\usepackage{float}
\usepackage{latexsym}
\usepackage{amsmath}
\usepackage{graphics}
\usepackage{amssymb}
\usepackage{layout}
\usepackage{verbatim}
\usepackage{amsfonts,epsfig}
\usepackage{color}
\usepackage{setspace}
\newcommand{\beqnar}{\begin{eqnarray}}
\newcommand{\eeqnar}{\end{eqnarray}}

\newcommand{\bk}{{\bf k }}

\newcommand{\bq}{{\bf q }}

\newcommand{\beq}{\begin{equation}}
\newcommand{\eeq}{\end{equation}}

\begin{document}
\title{Finite temperature inelastic mean free path and quasiparticle lifetime in graphene}
\author{Qiuzi Li and S. Das Sarma}
\affiliation{Condensed Matter Theory Center, Department of Physics, University of Maryland, College Park, Maryland 20742}
\date{\today}
\begin{abstract}
We adopt the GW approximation and random phase approximation to study finite temperature effects on the inelastic mean free path and quasiparticle lifetime by directly calculating the imaginary part of the finite temperature self-energy induced by electron-electron interaction in extrinsic and intrinsic graphene. In particular, we provide the density-dependent leading order temperature correction to the inelastic scattering rate for both single-layer and double-layer graphene systems. We find that the inelastic mean free path is strongly influenced by finite-temperature effects. We present the similarity and the difference between graphene with linear chiral band dispersion and conventional two dimensional electron systems with parabolic band dispersion. We also compare the calculated finite temperature inelastic scattering length with the elastic scattering length due to  Coulomb disorder, and comment on the prospects for quantum interference effects showing up in low density graphene transport. We also carry out inelastic scattering calculation for electron-phonon interaction, which by itself gives rather long carrier mean free paths and lifetimes since the deformation potential coupling is weak in graphene, and therefore electron-phonon interaction contributes significantly to the inelastic scattering only at relatively high temperatures.
\end{abstract}


\maketitle

\section{Introduction}

Ever since the pioneering work by Novoselov and Geim\cite{Novoselov}, graphene has attracted attention because of its potential application in future nanotechnology. Its chiral 2D linear dispersion leads to novel properties,  distinguishing it from conventional parabolic band semiconductors. Electron-electron ($e$-$e$) interaction in graphene is of great interest, both experimentally\cite{Bostwick_NatPhy07,SYZhou_PRB08,Kozikov_PRB10,Elias_NatPhy11} and theoretically\cite{DasHwangTse_PRB07,hwang_PhysicaE08,Polini_PRB08,Shutt_PRB11,dassarma2010,Kotov_RMP12}, because it plays an important role in determining electronic properties of graphene. The reduced dimensionality dramatically enhances the electron-electron interaction effects\cite{altshuler} while chirality and linear dispersion have subtle qualitative and quantitative effects distinguishing graphene from the usual non-chiral parabolic semiconductor-based 2D systems.

Experimentally, the inelastic mean free path of graphene due to $e$-$e$ interaction, denoted as $l$ throughout this paper, is an essential parameter relevant for possible ultra-fast device applications because this length  defines the distance over which the carriers move without any energy loss\cite{Tse_APL08}. In addition, several experimental groups have investigated the temperature and gate-voltage dependence of phase coherence length in graphene\cite{DKKi_PRB08,Tikhonenko_PRL08,Tikhonenko_PRL09}, which is directly determined by $l$, although may not always be identical to it.  Thus, $l$ plays an important role in determining quantum interference induced localization phenomenon.  The electro-electron interaction induced inelastic scattering  is a dominant factor in determining the weak localization effects at low temperatures\cite{DKKi_PRB08} since the phase coherence length mainly originates from Coulomb interaction among electrons. In particular, the inelastic mean free path or more specifically whether it is longer or shorter than the elastic transport mean free path is an important ingredient in understanding the origin of the recently observed metal-insulator transition in high-purity graphene devices\cite{Ponomarenko_arX11,DasHwangLi_PRB12,kostya}.

Theoretically, the inelastic quasiparticle lifetime of extrinsic or doped graphene  at zero temperature has been considered in the literature\cite{Gonzalez_PRL96,HwangHuInelastic_PRB07}. (In intrinsic undoped graphene, the inelastic quasiparticle lifetime due to $e$-$e$ interaction vanishes at the Dirac point, indicating the Dirac point to be a non-Fermi-liquid unstable point\cite{DasHwangTse_PRB07}, but this is not a problem at any finite carrier density, which is the case we primarily study in this paper.) In doped extrinsic graphene, it has been found that the chiral linear dispersion of graphene results in qualitative differences in the energy dependence of inelastic quasiparticle scattering rates compared with conventional parabolic two dimensional electron systems (2DES). Recently, Sch\"{u}tt {\it et al.} have analyzed the inelastic scattering rate induced by Coulomb interaction for intrinsic, i.e. undoped graphene\cite{Shutt_PRB11}. They have shown that finite temperature strongly affects the inelastic quasiparticle lifetime at the Dirac point. The inelastic scattering rate corresponds to the so-called on-shell imaginary part of self energy $\text{Im}\Sigma^R(\bk,\xi_{\bk})$, $\bk$ being the momentum and $\xi_{\bk}$ the quasiparticle energy measured from the chemical potential $\mu$. So far, neither the inelastic mean free path nor the imaginary part of the self-energy of extrinsic graphene at finite temperature has been explicitly calculated. In addition, the effect of finite temperature on the off-shell imaginary part of the self-energy of intrinsic graphene has not yet been studied.

In this paper, we theoretically study the inelastic mean free path $l$ and the imaginary part of the retarded self-energy $\text{Im} \Sigma^R(\bk,\omega)$ in graphene within the leading order many-body perturbative GW approximation. The GW approximation involves the leading-order theory in the dynamically screened Coulomb interaction which includes all the ring diagrams in the electron self-energy.  We generalize the previous GW  work\cite{JWMacDo_PRB96,ZhengDas_PRB96,HwangHuInelastic_PRB07} in the literature (carried out at $T=0$) to  finite temperatures, for both extrinsic and intrinsic graphene. We find that the finite temperature results for $l$ and $\text{Im} \Sigma^R (\bk,\omega)$ are very different from the zero temperature ones due to a dramatic change in the dynamic dielectric function with increasing temperature.  We also provide an analytical expression for the on-shell imaginary part of the self-energy for a double-layer graphene system, which is relevant for a recent experiment\cite{Ponomarenko_arX11}. We also calculate the imaginary part of the electron self-energy due to electron-phonon ($e$-$p$) interaction in graphene through the deformation potential coupling, and provide results for the corresponding inelastic mean free path as a function of temperature.  In general, phonon effects are rather small in graphene because the deformation potential coupling is very weak in graphene.  Thus, except at very high temperatures, phonon effects are negligible in the determination of the inelastic mean free path and quasiparticle lifetime.

The rest of this paper is organized as follows. In Sec.~\ref{sec:theory}, we introduce the theoretical formalism for treating inelastic $e$-$e$ scattering at finite temperatures in  both monolayer and  double-layer graphene systems. We also obtain analytical expressions for the imaginary part of the retarded self-energy in the low energy and low temperature limit. In Sec.~\ref{sec:numerical} we provide numerical calculations of the imaginary part of the self-energy (which is the inverse of the quasiparticle lifetime) and the associated inelastic mean free path as a function of carrier energy and temperature. We also compare the inelastic mean free path with the disorder induced elastic mean free path. In Sec.~\ref{sec:eph}, we provide the electron-phonon interaction results separately, giving both the basic theory briefly and the numerical results. In Sec.~\ref{sec:conclu}, we discuss and summarize the main results of this paper.

\section{Theoretical formalism}
\label{sec:theory}

In this section, we provide the theoretical formalism for evaluating the finite temperature imaginary part of the retarded self-energy due to $e$-$e$ interaction in graphene.  We also obtain the asymptotic behavior of  the imaginary part of the self-energy in the low energy and low temperature limit.

\subsection{Imaginary part of self-energy for extrinsic doped graphene}
The imaginary part of the retarded self-energy of monolayer graphene within GW approximation can be expressed as\cite{DasHwangTse_PRB07,Tse_APL08}:
\begin{eqnarray}
\text{Im} \Sigma_{s}^{R}(k, \omega) &=& -\frac{1}{2}\sum_{\bq,s'=\pm}V_q [n_B(\xi_{\bk+\bq,s'}-\omega)+n_F(\xi_{\bk+\bq,s'})]
\nonumber  \\
&& (1+s s' \cos\theta)\text{Im}\left[\frac{1}{\epsilon(q,\xi_{\bk+\bq,s'}-\omega)}\right]
\label{eq:exim}
\end{eqnarray}
where functions $n_F$ and $n_B$ denote the Fermi and Bose distribution functions, respectively. $V_q = 2 \pi e^2/(\kappa q)$ is the Coulomb interaction in  momentum space with average background dielectric constant $\kappa$.  $s, s' = \pm 1$ are band indices.  $\xi_{\bk+\bq,s} = s \varepsilon_{\bk+\bq}-\mu$ ($\varepsilon_{\bk}=\hbar v_F |\bk|$ with graphene Fermi velocity $v_F =10^6$ m/s) is the carrier energy relative to the finite temperature, non-interacting chemical potential $\mu$, determined by the conservation of the total electron density. $\epsilon(q,\omega) = 1 + V_q \Pi(q,\omega)$ is the finite temperature dynamic dielectric function within random-phase approximation (RPA)\cite{HwangDas_PRB07,ramezanali_JPMT09,hwangDrag_PRB11}, and $\Pi(q,\omega)$ is the irreducible polarizability. Note that  both single particle excitations ($\text{Im}[\epsilon] \neq 0$) and plasmon excitations ($\text{Re}[\epsilon] = 0$ and $|\text{Im}[\epsilon]| = 0^+$) contribute to $\text{Im} \Sigma_{s}^{R}(k, \omega)$. We use $\hbar = 1$ throughout.

The inverse quasiparticle lifetime or the quasiparticle scattering rate $1/\tau = \Gamma_s(\bk)$ is directly related to the imaginary part of the on-shell self-energy, i.e.,
\beq
\dfrac{1}{\tau} = \Gamma_s(\bk) = 2 \text{Im} \Sigma^R_s (\bk, \xi_{\bk, s}) 
\eeq

The inelastic mean free path $l$ due to $e$-$e$ interaction is then given by
\beq
l = v_F \tau
\label{eq:inelaee}
\eeq
where $v_F$ is the non-interacting Fermi velocity\cite{HwangHuInelastic_PRB07}. The central quantity in our calculation is the imaginary part of the self-energy which determines the quasiparticle lifetime.  The inelastic length is a derived quantity arising from the quasiparticle lifetime, not calculated directly by the theory. For consistency in staying within the leading order in dynamically screened Coulomb interaction\cite{ZhengDas_PRB96,JWMacDo_PRB96}, the velocity should not have any interaction dependence, and thus should be the bare Fermi velocity. Multiplying quansiparticle lifetime by renormalized velocity will mix different orders in the perturbation theory\cite{yingdas_PRB05,quinn_PR58,HwangHuInelastic_PRB07}. 

\subsubsection{Zero temperature analytical results $T=0$}
\label{subsubsec:ext0}
We first give the analytical formula for $\text{Im} \Sigma_{+}^{R}(k, \xi_k)$  at zero temperature and for $\varepsilon_F \gg \xi_k$. This  has already been given in Ref.~[\onlinecite{DasHwangTse_PRB07}] and we just provide the results below for completeness and comparison.

The quasiparticle scattering rate can be obtained through the imaginary part of the self-energy within the on-shell approximation $\omega = \xi_{\bk,s}$:
\begin{eqnarray}
&\text{Im} \Sigma_{+}^{R}(k, \xi_k)&= \frac{1}{2}\sum_{s'=\pm}\sum_{\bq}V_q[\vartheta(\varepsilon_k- s'\varepsilon_{\bk+\bq})-\vartheta(-\xi_{\bk+\bq,s'})]\nonumber \\
& &\times (1+s'\cos\theta)\text{Im}\left[\frac{1}{\epsilon(q,s'\varepsilon_{\bk+\bq}-\varepsilon_{\bk}+i 0^+)}\right]
\label{eq:im1}
\end{eqnarray}
where $\vartheta (x)$ is the Heaviside unit step function.

From Eq.~\eqref{eq:im1}, it is straightforward to see that the interband scattering from the valence band $s'=-1$ vanishes at zero temperature. In addition, the plasmon emission process also vanishes in calculating the on-shell imaginary part of the self-energy because the plasmon frequency requires $\omega_{pl}(q) > \hbar v_F q$ for extrinsic graphene at zero temperature\cite{HwangDas_PRB07} but $|\varepsilon_{\bk+\bq}-\varepsilon_{\bk}| < \hbar v_F q$ in the dynamic dielectric function. Thus, Eq.~\eqref{eq:im1} reduces to :
\begin{eqnarray}
\text{Im} \Sigma_{+}^{R}(k, \xi_k)=\frac{e^2}{4\pi \kappa} \int_0^{2\pi}d\theta (1+\cos\theta)\int_{k_F}^k dk' k' \nonumber \\
\times \frac{\text{Im}[1/\epsilon(|\bk-\bk'|,\varepsilon_{k'}-\varepsilon_k+i 0^+)]}{\sqrt{k^2+k'^2-2k k' \cos\theta}}
\label{eq:im2}
\end{eqnarray}

In the long-wavelength $x=q/2k_F \ll 1$ limit, we can obtain the analytical expression for Eq.~\eqref{eq:im1}. In this limit, the dominant contribution to $\text{Im}[1/\epsilon]$ comes from low energies $u=\omega/v_F q \ll 1$, where the irreducible polarizability in the leading order is given  by\cite{HwangDas_PRB07}
\beq
\Pi(q,\omega)\simeq D_0 [1+ i \frac{\omega}{v_F q}]
\eeq
where $D_0 = \frac{2 k_F}{\pi \hbar v_F}$, the density of states at Fermi energy.

Then, the asymptotic behavior of the  imaginary part of the dynamic dielectric function can be written as:
\beq
\text{Im}\dfrac{1}{\epsilon(q,\varepsilon_{\bk+\bq}-\varepsilon_k+i 0^+)}
\simeq -\dfrac{k'-k}{q_{TF}}
\label{eq:anadie}
\eeq
where $k'= |\bk+\bq|$ and $q_{TF} = \frac{4 k_F e^2}{\hbar v_F \kappa}$ denotes the Thomas-Fermi screening wave-vector in graphene. To obtain Eq.~\eqref{eq:anadie}, we have used the fact that the integrand in Eq.~\eqref{eq:im2} has sharp peaks near forward scattering  momentum transfer, i.e. $q=0$. The asymptotic expansion of the imaginary part of the self-energy in Eq.~\eqref{eq:im2} can then be obtained by integrating out the angular part and expanding in the small momentum transfer $k-k' \ll k_F$:
\beq
\text{Im} \Sigma_{k+}^{R}\simeq -\dfrac{\xi_k^2}{8\pi \varepsilon_F}\left[\ln\left(\dfrac{\xi_k}{8 \varepsilon_F}\right)+\dfrac{1}{2}\right]
\label{eq:anaT0}
\eeq
The leading order energy correction to the on-shell imaginary part of the self-energy is $\xi_k^2 \ln \xi_k$,  similar to that obtained for  conventional two-dimensional electron systems (2DES)\cite{Giuliani_PRB82}. The subleading term in Eq.~\eqref{eq:anaT0} has opposite sign to that in 2DES because of the chiral nature of graphene and the absence of interband contribution from valence band ($s'=-1$) at zero temperature.

\subsubsection{Finite temperature analytical results $T > 0$}
To maintain analytic tractability for $\text{Im} \Sigma_{k+}^{R}$ at finite temperature, we consider only low temperatures and small $\xi_k$, i.e., $\varepsilon_F \gg k_B T \gg \xi_k$. In this limit, both the interband scattering  and the temperature dependence of $\text{Im}[1/\epsilon]$, being exponentially suppressed at low temperatures, can be ignored. Applying the same technique as in the $T=0$ case, the finite temperature on-shell imaginary part of the self-energy can be approximated as:
\begin{eqnarray}
\text{Im} \Sigma_{k+}^{R}&=&\frac{e^2}{4\pi \kappa} \int_0^{2\pi}d\theta (1+\cos\theta)\int_{0}^{\infty} dk' k' [n_B(\varepsilon_{k'}-\varepsilon_k)\nonumber \\
&+&n_F(\varepsilon_k)]\frac{\text{Im}[1/\epsilon(|\bk-\bk'|,\varepsilon_{k'}-\varepsilon_k+i 0^+)]}{\sqrt{k^2+k'^2-2k k' \cos\theta}}
\label{eq:im3}
\end{eqnarray}
The imaginary part of the inverse dielectric function can still be approximated as in Eq.~\eqref{eq:anadie}. After integrating out the angular part and expanding in the limit $\xi_{k} \ll \varepsilon_F$, the momentum integration is very straightforward and we obtain an asymptotic formula for Eq.~\eqref{eq:im3} in the limit $\varepsilon_F \gg k_B T \gg \xi_k$:
\begin{eqnarray}
\text{Im} \Sigma_{k+}^{R}\simeq  -\dfrac{\pi(k_B T)^2 }{8 \varepsilon_F}\left[\ln\left(\dfrac{k_B T}{8 \varepsilon_F}\right)+1.08387\right]
\label{eq:anaT}
\end{eqnarray}
Thus, the leading order temperature correction to the on-shell $\text{Im} \Sigma_{k+}^{R}$ goes as $t^2 \ln t$ with $t= T/T_F$ ($T_F$ being the Fermi temperature).

\subsection{Imaginary part of self-energy for double-layer graphene}

In this subsection, we provide analytical results for $\text{Im} \Sigma_{k+}^{R}$ of a double-layer graphene system consisting of a ``studied" and a ``control" layer of graphene separated by an insulating layer\cite{Ponomarenko_arX11}, which has recently been studied experimentally. These two layers of graphene are electronically isolated but the screening effect of the control layer must be considered. We will only focus on the doped double-layer case, where Fermi levels of both layers are above the Dirac point.

The imaginary part of the self-energy of the ``studied" layer for a double-layer graphene, in the presence of screening by carriers both in the control layer and the studied layer itself, can be written as\cite{liandas_PRB1996,LianMac_PRB94}:
\begin{eqnarray}
\text{Im} \Sigma^R_{s}(k, \omega) &=&  \frac{1}{2}\sum_{s'=\pm,\bq}[n_B(\xi_{\bk+\bq,s'}-\omega)+n_F(\xi_{\bk+\bq,s'})] V (q)\nonumber
\\
&\times & (1+s s' \cos\theta) \text{Im}
\left[\frac{1}{\epsilon^{scr} (q,\xi_{\bk+\bq,s'}-\omega)}\right]
\label{eq:dblim1}
\end{eqnarray}
$\epsilon^{scr} (q,\xi_{\bk+\bq,s'}-\omega)$ is the RPA dynamic dielectric function for the double-layer system, which incorporates screening effects from a nearby ``control" graphene layer in addition to the usual screening by the studied layer electrons themselves\cite{liandas_PRB1996,LianMac_PRB94,profumo_PRB10}.
\beq
\frac{1}{\epsilon^{scr} (q,\omega)}= \frac{1 + V(q) \Pi_{22} (1- e^{-2q d})}{1+ V(q) [\Pi_{11}+\Pi_{22}]+ \Pi_{11}\Pi_{22}V(q)^2[1- e^{-2q d}]}
\label{eq:dblscr}
\eeq
where $\Pi_{11}$ and $\Pi_{22}$ are polarization operators of the studied and the control layer, respectively.  $d$ is the separation between the two layers. Note that Eq.~\eqref{eq:dblscr} reduces to the dynamic dielectric function of monolayer graphene in the large $d$ limit as it should because the two layers become decoupled in the large $d$ limit.

\subsubsection{Zero temperature analytical results $T=0$}
In this section, we derive the analytical formula for the imaginary part of the double-layer self-energy at zero temperature in the limit $\varepsilon_F \gg \xi_k$. This is very close to what we derived in the previous subsection for monolayer graphene but with different prefactors coming from the additional screening effects of the second nearby  layer.

At zero temperature, the interband scattering from the valence band $s'=-1$ vanishes in Eq.~\eqref{eq:dblim1}, and the on-shell expression reduces to :
\begin{eqnarray}
\text{Im} \Sigma_{k+}^{R}=\frac{e^2}{4\pi \kappa} \int_0^{2\pi}d\theta (1+\cos\theta)\int_{k_F}^k dk' k' \nonumber \\
 \frac{\text{Im}[1/\epsilon^{scr}(|\bk-\bk'|,\varepsilon_{k'}-\varepsilon_k+i 0^+)]}{\sqrt{k^2+k'^2-2k k' \cos\theta}}
\label{eq:dblim2}
\end{eqnarray}

In the long-wavelength $x=q/2k_{F (1,2)} \ll 1$ and low energy  limit $u=\omega/v_F q \ll 1$, it is possible to perform an analytical evaluation of Eq.~\eqref{eq:dblim2}. In this case, the irreducible polarizability of graphene is given in the leading order by\cite{HwangDas_PRB07}
\begin{eqnarray}
\Pi_{11}(q,\omega)\simeq D_{011} [1+ i \frac{\omega}{v_F q}]
\\
\Pi_{22}(q,\omega)\simeq D_{022} [1+ i \frac{\omega}{v_F q}]
\end{eqnarray}
where $D_{011} = \frac{2 k_{F1}}{\pi \hbar v_F}$ and $D_{022} = \frac{2 k_{F2}}{\pi \hbar v_F}$ correspond to the density of states at Fermi energy of the studied and the control layer, respectively.

Then, the imaginary part of the screened dielectric function is approximated as:
\begin{eqnarray}
&&\text{Im}\dfrac{1}{\epsilon^{scr}(q,\varepsilon_{\bk+\bq}-\varepsilon_k+i 0^+)} \simeq -(k'-k)
\nonumber
\\
&&\times \dfrac{q_{TF1}+q_{TF2}+ 4q_{TF1} q_{TF2}d +4q_{TF1} ( q_{TF2}d)^2}{(q_{TF1}+q_{TF2}+ 2q_{TF1} q_{TF2}d )^2}
\label{eq:dbldielec}
\end{eqnarray}
where $q_{TF1} = \frac{4 k_{F1} e^2}{\hbar v_F \kappa}$ and $q_{TF2} = \frac{4 k_{F2} e^2}{\hbar v_F  \kappa}$ are the Thomas Fermi wave-vectors of the studied and the control graphene layer. In deriving Eq.~\eqref{eq:dbldielec}, we have used the long-wave length limit $ |\bk-\bk'|/2k_F \ll 1$, which also indicates that $|\bk-\bk'| d \ll 1 $ for $d \sim O (1 \ \text{nm})$  and $\frac{(k'-k)}{|\bk-\bk'|}  \ll 1$. From Eq.~\eqref{eq:dbldielec}, we see that the difference in the analytical expressions for $\text{Im}\Sigma^R_+ $ between monolayer and double-layer graphene comes from the second line of Eq.~\eqref{eq:dbldielec}.

Using the approximation of Eq.~\eqref{eq:dbldielec} in Eq.~\eqref{eq:dblim2}, we obtain the final analytical formula:
\begin{eqnarray}
&&\text{Im} \Sigma_{k+}^{R}
\simeq  -\dfrac{\xi_k^2 }{8\pi \varepsilon_{F1}} \left[\ln\left(\dfrac{\xi_k}{8 \varepsilon_{F1}}\right)+\dfrac{1}{2}\right]\nonumber \\
&&\times  \dfrac{1+ \sqrt{n_2}/\sqrt{n_1} + 16 r_{s} d \sqrt{\pi n_2} + 64 r_{s}^2 d^2 \pi n_2}{ (1+ \sqrt{n_2}/\sqrt{n_1} + 8 r_{s} d \sqrt{\pi n_2})^2}
\label{eq:blgt0}
\end{eqnarray}
where $n_1$, $n_2$ are the carrier densities of the studied layer and the control layer respectively. $r_{s} = e^2 / (\hbar v_F \kappa)$ is the graphene fine structure constant defining the electron-electron interaction strength and $\varepsilon_{F1}$ is the Fermi energy of the studied layer. It is clear that  Eq.~\eqref{eq:blgt0}, as expected,  reduces to the monolayer graphene result as given in Eq.~\eqref{eq:anaT0} for two extremes: $n_2 = 0$ and $d \rightarrow \infty$.

\subsubsection{Finite temperature analytical results $T > 0$}
Next, we provide the analytical expression for the imaginary part of the self-energy of a double-layer graphene in the low $T$ and small $\xi$ limit, i.e., $\varepsilon_{F(1,2)} \gg k_B T \gg \xi_k$ as before.

In the low temperature limit, we neglect both the interband scattering and the temperature-dependent imaginary part of the dielectric function, which are exponentially suppressed. Then, the asymptotic expansion of Eq.~\eqref{eq:dblim1} is given by:
\begin{eqnarray}
&&\text{Im} \Sigma_{k+}^{R} \simeq -\dfrac{\pi(k_B T)^2 }{8 \varepsilon_{F1}} \left[\ln\left(\dfrac{k_B T}{8 \varepsilon_{F1}}\right)+1.08387\right]\nonumber
\\
&&\times \dfrac{1+ \sqrt{n_2}/\sqrt{n_1} + 16 r_{s} d \sqrt{\pi n_2} + 64 r_{s}^2 d^2 \pi n_2}{ (1+ \sqrt{n_2}/\sqrt{n_1} + 8 r_{s} d \sqrt{\pi n_2})^2}
\end{eqnarray}

\subsection{Imaginary part of self-energy for intrinsic graphene}
The imaginary part of the self-energy for undoped or intrinsic monolayer graphene can be obtained by setting $\mu \equiv 0$ in Eq.~\eqref{eq:exim}, in which case the chemical potential is always independent of temperature.

At zero temperature, the analytical form of the irreducible polarizability for intrinsic graphene \cite{HwangDas_PRB07} is:
\beq
\Pi = \dfrac{q^2}{4} \left( \dfrac{\theta(v_F q - \omega)}{\sqrt{(v_F^2 q^2 - \omega^2)}} + i \dfrac{\theta( \omega- v_F q)}{\sqrt{( \omega^2- v_F^2 q^2)}}\right)
\label{eq:intpola}
\eeq
Then, the imaginary part of the self-energy within the on-shell approximation $\varepsilon = \varepsilon_k$ at zero temperature reduces to the product of two $\vartheta$ functions \cite{DasHwangTse_PRB07}:
\beq
\text{Im} \Sigma_{+}^{R}(k, \varepsilon_k) \sim \Sigma_{\bq} \vartheta(|\varepsilon_{\bk+\bq}-\varepsilon_{\bk}|-\varepsilon_q)\vartheta(\varepsilon_{\bk}-\varepsilon_{\bk+\bq})
\label{eq:inim1}
\eeq
The first $\vartheta$ function is due to the imaginary part of the dynamic dielectric function while the second one comes from the sum of the Fermi and Bose distribution functions at $T=0$. It is easy to see from the above equation that $\text{Im} \Sigma_{+}^{R}(k, \varepsilon_k)$ vanishes because of phase space restrictions imposed by the $\vartheta$ function. On the other hand, we can evaluate the zero temperature imaginary part of the self-energy at zero momentum $k =0$ and finite energy $\omega$ by using the  polarizability given in Eq.~\eqref{eq:intpola}.

Then the imaginary part of the self-energy at $k \equiv 0 $ is given by:
\beq
\text{Im} \Sigma_{\pm}^{R}(k=0, \omega) =\omega f(r_s)
\label{eq:intt0}
\eeq
with
\beq
f(x) = \dfrac{2}{\pi^2 x} \left( \pi (1-x) + \dfrac{2- (\pi x/2)^2}{\sqrt{(\pi x)^2 -4}} \ln \dfrac{\pi x-\sqrt{(\pi x)^2 -4}}{\pi x+\sqrt{(\pi x)^2 -4}}\right)
\eeq
which was already obtained in Ref.~[\onlinecite{DasHwangTse_PRB07}].  Note that for the general case $(\bk, \omega)$, it is almost impossible to get an analytical formula for the imaginary part of the self-energy due to the complicated angular integration. However, we know that $\text{Im} \Sigma_{+}^{R} (\bk, \omega) \equiv 0 $ for $\omega < \hbar v_F k$ at zero temperature because of the combination of the two $\vartheta$ functions mentioned above.

Next, we mention that the analytical formula for the finite temperature imaginary part of the self-energy for the intrinsic undoped graphene is quite tricky necessitating a  very careful analysis of the finite temperature dynamic dielectric function. Since undoped graphene is not the focus of our current work, we refer the reader to Ref.~[\onlinecite{Shutt_PRB11}], where the  on-shell imaginary part of the self-energy in intrinsic graphene has been recently studied. Note that the finite temperature imaginary part of the on-shell self-energy is proportional to the temperature $T$, as can be shown by dimensional counting and can also be seen from the numerical results shown in Sec.~\ref{sec:numerical}. In addition, we find that the leading order temperature correction to the imaginary part of the self-energy for two cases: $k=0$ and $k_B T \ll |\omega - \varepsilon_{\bk}|$, are exponentially suppressed. Since the self-energy for the undoped case has already been studied analytically in some details in Ref.~[\onlinecite{Shutt_PRB11}] we do not provide  any further discussion of this issue in this paper.  We also add that from the experimental perspective, pure intrinsic graphene is uninteresting since it is unstable to any (for example, disorder-induced) density fluctuations which would locally dope the Dirac point, and thus, all experimentally studied graphene samples are likely to be extrinsic doped graphene.

\section{Numerical Results}
\label{sec:numerical}
In this section, we provide our numerical results for both inelastic mean free  path and the imaginary part of the self-energy (which is essentially the inelastic scattering rate or the inverse of the quasiparticle lifetime) in graphene for both extrinsic graphene and intrinsic graphene. If not specified, the effective background dielectric constant used in our calculation is $\kappa = 5$, i.e. $r_s = 0.44$, corresponding to monolayer graphene sandwiched between two Boron-Nitride layers\cite{Ponomarenko_arX11}. We compare our numerical results with the asymptotic analytical results derived in Sec.~\ref{sec:theory}. We also calculate the ratio between the inelastic mean free path and the elastic mean free path induced by long-range Coulomb disorder as a function of temperature in order to assess the importance of possible quantum interference effects in highly resistive graphene samples at low temperatures.

\subsection{Inelastic mean free path and imaginary part of self-energy for extrinsic graphene}

\begin{figure}
\begin{center}
\includegraphics[width=0.99\columnwidth]{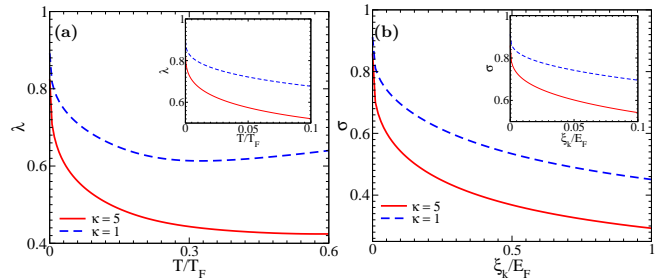}
\caption{(Color online).  (a). Calculated  ratio $\lambda$, which is the numerical result of $\text{Im}\Sigma^R_+(\bk_F, \xi_{\bk_F})$  divided by the analytical asymptotic formula $-\dfrac{\pi(k_B T)^2 }{8 \varepsilon_F}\left[\ln\left(\dfrac{k_B T}{8 \varepsilon_F}\right)+1.08387\right]$ (given in Eq.~\eqref{eq:anaT}), as function of $T/T_F$. The inset presents $\lambda$ in the low temperature regime. (b) Calculated ratio $\sigma$, the numerical result of  $\text{Im}\Sigma^R_+(\bk, \xi_{\bk})$ divided by $-\dfrac{\xi_k^2}{8\pi \varepsilon_F}\left[\ln\left(\dfrac{\xi_k}{8 \varepsilon_F}\right)+\dfrac{1}{2}\right]$ (given in Eq.~\eqref{eq:anaT0}) as a function of $\xi_{\bk}/E_F$ at $T=0$. The inset presents $\sigma$ in the low energy regime. The solid and dashed lines correspond to $\kappa = 5$ ($r_s = 0.44$) and $\kappa = 1$ ($r_s = 2.2$), respectively.
}
\label{fig:norm}
\end{center}
\end{figure}

In Fig.~\ref{fig:norm}, we compare our calculated analytical results of  the on-shell imaginary part of the self-energy with the numerical results, which show good agreement with each other. The ratio $\lambda$ is $\text{Im}\Sigma^R_+(\bk_F, \xi_{\bk_F})$, calculated numerically at finite temperature,  divided by its value given by Eq.~\eqref{eq:anaT}, while the ratio $\sigma$ is  $\text{Im}\Sigma^R_+(\bk, \xi_{\bk})$, calculated numerically at zero temperature, divided by its value given by Eq.~\eqref{eq:anaT0}. Both ratios $\lambda$ and $\sigma$ are closer to unity for smaller background dielectric constant $\kappa$. This is because the approximation used in  Eq.~\eqref{eq:anadie} is more accurate for a larger value of $q_{TF}$. In Fig.~\ref{fig:norm}(a), the non-monotonic behavior in $\lambda$ for $\kappa = 1$ is due to its denominator, i.e., Eq.~\eqref{eq:anaT}, which is a non-monotonic function of $T/T_F$. For $\kappa = 5$, the non-monotonic behavior in $\lambda$ will show up at higher $T/T_F$. Similarly, the non-monotonic behavior of $\sigma$ will show up at larger values of $\xi_k/E_F$. Note that both ratios $\lambda$ and $\sigma$ are independent of Fermi-energy $E_F$ for a fixed fine structure constant $r_s$ within the GW approximation. We want to mention that both ratios in graphene have opposite trends compared with 2DES in the sense that in the latter system $\lambda$ and $\sigma$ are both larger than unity as they approach unity asymptotically\cite{ZhengDas_PRB96,JWMacDo_PRB96} whereas in graphene the ratios always lie below unity.  We mention that it is quite interesting to note that the asymptotic expressions, although they are derived for energy or temperature being much smaller than $E_F$ or $T_F$, appear to be valid quantitatively within a factor of two well outside the asymptotic regime, i.e. even when $T \sim T_F$.

\begin{figure}
\begin{center}
\includegraphics[width=0.99\columnwidth]{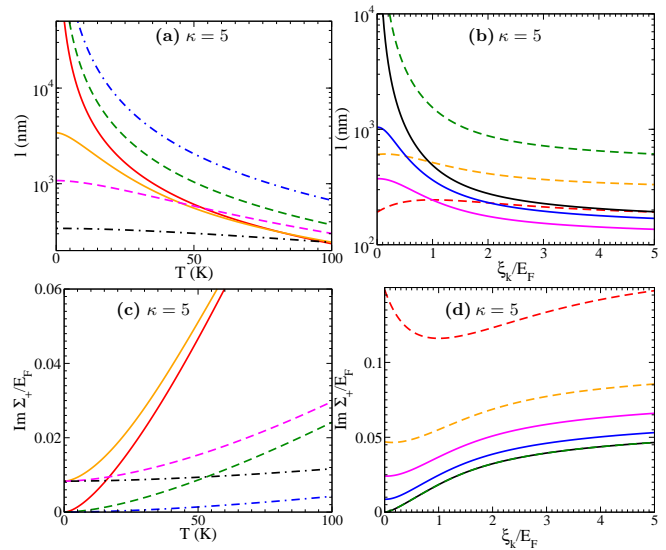}
\caption{(Color online).  Calculated inelastic scattering length $l$ and the associated on-shell imaginary part of the self-energy $\text{Im} \Sigma_{+}^{R}(k, \xi_k)/E_F$ for extrinsic graphene  ($\varepsilon_F > 0$) with dielectric constant $\kappa = 5$ ($r_s = 0.44$). (a) $l$ as a function of temperature $T$.  The solid, dashed and dot-dashed  lines are for carrier density $n= 10^{10}$, $10^{11}$ and $10^{12}$ cm$^{-2}$, respectively. The upper three lines are for $\xi_k=\xi_{k_F}$ while the lower three lines are for $\xi_k=0.5\epsilon_{F}$. (b) $l$ as a function of $\xi_k/E_F$ for different temperatures. The solid and dashed lines are for carrier density $n= 10^{11}$, and $10^{10}$ cm$^{-2}$, respectively. For the same carrier density, the lines from top to bottom correspond to  $T= 0$ K, $50$ K, and $100$ K. (c) and (d) are the associated $\text{Im} \Sigma_{+}^{R}(k, \xi_{\bk})/E_F$ corresponding to (a) and (b), respectively.
}
\label{fig:eximl}
\end{center}
\end{figure}

Fig.~\ref{fig:eximl} shows our numerical results for the energy and temperature dependence of the inelastic mean free path $l$ and the associated on-shell imaginary part of the self-energy. In Fig.~\ref{fig:eximl}(a), we present the inelastic mean free path $l$ as a function of temperature for two different carrier energy values. We can see that $l$ is a monotonically  decreasing function of temperature. In particular, the injected electron can only decay via intraband processes at zero temperature as discussed in Sec.~\ref{subsubsec:ext0}. In addition, the decay via plasmon emission is  inhibited at zero temperature due to phase-space restrictions\cite{HwangHuInelastic_PRB07}. On the other hand, at finite temperatures, the injected electron can decay via both interband and intraband excitations and the region of single particle excitations also increases due to thermal smearing effects.  Fig.~\ref{fig:eximl}(b) shows the energy dependence of $l$ for different temperatures and carrier densities. At low temperatures, the mean free path of hot-electrons ($\epsilon_k> E_F$) is shorter than that of the quasiparticle in the vicinity of the Fermi energy.  However, very  interestingly, the energy dependence of $l$ becomes non-monotonic at higher temperatures (see the lowest dashed line in Fig.~\ref{fig:eximl}(b)). This arises from the competition between  damping through valence band and conduction band, which have opposite trends as a function of carrier energy. To be more specific, the scattering rate from the valence band contribution ($s'= -1$) is a monotonically decreasing function of $\xi_{\bk}/E_F$ while the contribution from the conduction band ($s'=1$) is a monotonically increasing function of $\xi_{\bk}/E_F$.   Figs.~\ref{fig:eximl}(c) and (d) provide the on-shell $\text{Im} \Sigma^R_+ (\bk,\xi_{\bk})$. We point out that $\text{Im}\Sigma^R_+(\bk,\xi_{\bk})/E_F$ for the two carrier densities at zero temperature in Fig.~\ref{fig:eximl}(d) are essentially the same  because the rescaled $\text{Im} \Sigma^R_+ (\bk/k_F,\xi_{\bk}/E_F)/E_F$ is universal at $T=0$. Notice that in our theoretical GW  formalism, we exclude multiparticle excitations, which have a relatively small effect on $l$ in graphene\cite{HwangHuInelastic_PRB07} because of the small values of $r_s$ implying weak $e$-$e$ interaction strength.

\begin{figure}
\begin{center}
\includegraphics[width=0.99\columnwidth]{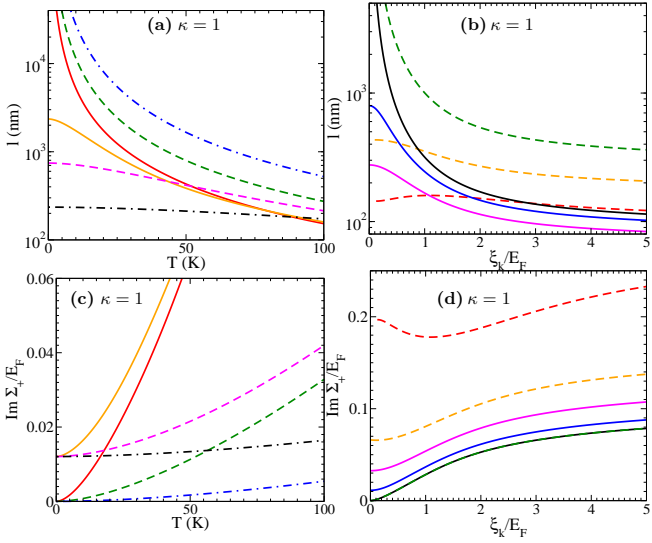}
\caption{(Color online).  Calculated inelastic scattering length $l$ and the associated on-shell imaginary part of the self-energy $\text{Im} \Sigma_{+}^{R}(k, \xi_k)/E_F$ for extrinsic graphene  ($\varepsilon_F > 0$) with dielectric constant $\kappa = 1$ ($r_s = 2.2$). (a) $l$ as a function of temperature $T$.  The solid, dashed and dot-dashed  lines are for carrier density $n= 10^{10}$, $10^{11}$ and $10^{12}$ cm$^{-2}$, respectively. The upper three lines are for $\xi_k=\xi_{k_F}$ while the lower three lines are for $\xi_k=0.5\epsilon_{F}$. (b) $l$ as a function of $\xi_k/E_F$ for different temperatures. The solid and dashed lines are for carrier density $n= 10^{11}$, and $10^{10}$ cm$^{-2}$, respectively. For the same carrier density, the lines from top to bottom correspond to  $T= 0$ K, $50$ K, and $100$ K. (c) and (d) are the associated $\text{Im} \Sigma_{+}^{R}(k, \xi_{\bk})/E_F$ corresponding to (a) and (b), respectively.
}
\label{fig:eximl_vac}
\end{center}
\end{figure}

For the purpose of comparison, in Fig.~\ref{fig:eximl_vac}, we show results of $l$ and $\text{Im}\Sigma_+$ for suspended graphene with the background dielectric constant $\kappa =1$ ($r_s = 2.2$). From Figs.~\ref{fig:eximl} and ~\ref{fig:eximl_vac}, we can see that the results of the  mean free path $l$ for different values of $r_s$  are qualitatively very similar. The inelastic scattering length $l$ is larger for smaller $r_s$ because of weaker electron-electron interaction. At low temperatures ($T<10$ K), our calculated quasiparticle mean free path for $\xi_k=\xi_{k_F}$ is on the order of $5 \times 10^3-10^5$ nm. However, the typical mean free path measured in the experiment will saturate at low temperatures and it is about a few  $\mu$m at liquid-helium temperature\cite{Tikhonenko_PRL08,Tikhonenko_PRL09}. This saturation is due to the sample size and this issue has been discussed by Tikhonenko et al. in Ref.~[\onlinecite{Tikhonenko_PRL08}].

\begin{figure}
\begin{center}
\includegraphics[width=0.99\columnwidth]{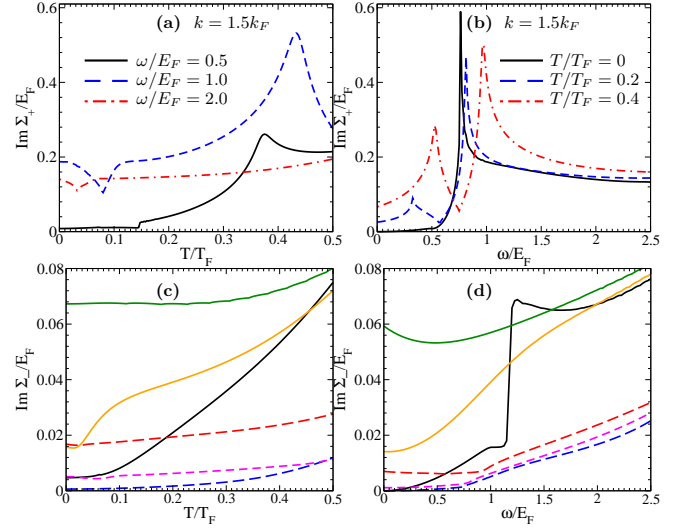}
\caption{(Color online).  Calculated Im$\Sigma^{R}(k, \omega)$ for extrinsic graphene (the chemical potential $\mu > 0$) with $\kappa = 5$. (a) and (b) correspond to conduction band while (c) and (d) correspond to valence band. In Fig. (a), the solid, dashed and dot-dashed lines correspond to $\omega/E_F = 0.5$, $1.0$ and $2.0$, respectively. (b) The solid, dashed and dot-dashed lines correspond to $T/T_F = 0$, $0.2$ and $0.4$, respectively. In Figs. (c) and (d), the solid and dashed lines correspond to $k=0$ and $k=1.5 k_F$. (c) For the same momentum, the lines from bottom to top correspond to $\omega/E_F = 0.5$, $1.0$ and $2.0$, respectively. (d) For the same momentum, the lines from bottom to top correspond to $T/T_F = 0$, $0.2$ and $0.4$, respectively. Note that Im$\Sigma^R_{+}(k=0, \omega) =\text{Im} \Sigma^R_{-}(k=0, \omega) $.
}
\label{fig:exim}
\end{center}
\end{figure}

To capture the effects of plasmons and interband single-particle excitations, we have to consider the off-shell imaginary part of the self-energy $\text{Im} \Sigma^R (\bk, \omega)$ ($\omega \neq \xi_{\bk,s}$), which is also needed to explain the experimental ARPES data\cite{Hwang_PRB081412,HwangHuInelastic_PRB07,hwang_PhysicaE08}. The numerical results for $\text{Im} \Sigma^R (\bk, \omega)$  are shown in Fig.~\ref{fig:exim}. Unlike the calculation for the on-shell imaginary part of the self energy,  we can see that the off-shell imaginary part of the self-energy contains two kinds of contributions from Eq.~\eqref{eq:im1}. The first comes from the single particle excitation occurring at $\text{Im} \epsilon \neq 0$. This can be further divided into intraband and interband excitations because of the gaplessness of graphene. The second contribution comes from plasmon excitations, where both $\text{Im} \epsilon$ and $\text{Re} \ \epsilon$ vanish. The plasmon resonance consists of a $\delta$ function at zero temperature, of which the width will be broadened as  temperature increases. Figs.~\ref{fig:exim} (a) and (b) present the rescaled conduction band $\text{Im} \Sigma^R_+ (\bk, \omega)/E_F$ with $k= 1.5 k_F$ as a function of rescaled temperature $T/T_F$ and energy $\omega/E_F$. We point out that $\text{Im} \Sigma^R_+ (\bk/k_F, \omega/E_F)/E_F$ is universal, not depending on the position of the Fermi level but only on the graphene fine structure constant $r_s$. The non-monotonic temperature dependence of $\text{Im} \Sigma^R_+$ for fixed $\omega/E_F$ as shown in Fig.~\ref{fig:exim}(a) comes from the competition among interband, intraband and plasmon excitations. In the zero temperature limit, it has been shown in Ref.~[\onlinecite{HwangHuInelastic_PRB07}] that the plasmon contribution dominates in the low energy regime for $k/k_F \gtrsim 1$ and it vanishes in the higher energy regime.  As the temperature goes up, the plasmon contribution is not a $\delta$ function anymore and it  gets broadened by Landau damping (see Fig.~\ref{fig:exim} (b)). Also, the boundaries of both inter- and intra-band excitations are thermally smeared. The above two effects give rise to the appearance of a smaller bump in $\text{Im} \Sigma^R_+$ in the lower energy region at finite temperature as shown in Fig.~\ref{fig:exim} (b). Figs.~\ref{fig:exim}(c) and (d) show the imaginary part of the self-energy for the valence band $\text{Im} \Sigma^R_{-}$.  $\text{Im} \Sigma^R_{-}$ is generally an increasing function of temperature.  In the $k=0$ and $T=0$ limit, only single particle excitations contribute to $\text{Im} \Sigma^R_{-} (0,\omega)$ and there is no plasmon emission because it requires $q < k_F$. But the integrand in Eq.~\eqref{eq:exim} has non-zero values only for $q> k_F$, being restricted by the sum of the Bose and Fermi distribution functions, i.e., $(\omega + E_F - \varepsilon_q)(E_F - \varepsilon_q)<0$. In particular,  only intraband single particle excitations contribute for energy $\omega/E_F \lesssim 1$ and the interband single particle excitation contribution increases sharply around $\omega/E_F \sim 1$. On the other hand, for $k=0$ but at nonzero temperature, $\text{Im} \Sigma^R (0,\omega)$ becomes much smoother compared with the zero temperature case. In addition, $\text{Im} \Sigma^R (0,0)$ has a finite value increasing with temperatures due to the thermally smeared boundaries of single particle excitation continua. Note that $\text{Im} \Sigma^R_{-} = \text{Im} \Sigma^R_{+}$ for $k=0$ as seen from Eq.~\eqref{eq:exim}, in which case the equation has no band-index dependence.

\subsection{Comparison between inelastic and elastic mean free path in extrinsic graphene}
\label{subsec:comparison}
\begin{figure}
\begin{center}
\includegraphics[width=0.9\columnwidth]{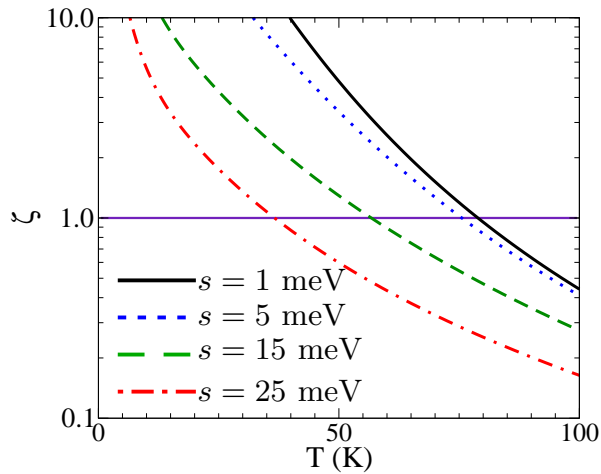}
\caption{(Color online).  Calculated ratio $\zeta = l/l_e$, the inelastic mean free path divided by the elastic mean free path, as a function of temperature for carrier density $n= 10^{9}$ cm$^{-2}$, $\kappa=5$, and charged impurity density $9 \times 10^{10}$ cm$^{-2}$. The solid, dotted, dashed and dot-dashed lines correspond to $s=1.0$, $5.0$, $15.0$ and $25.0$ meV, respectively, indicating increasing effects of Coulomb disorder induced inhomogeneous puddles.  The horizontal line indicates where (above the line) quantum interference could play a role.
}
\label{fig:compa}
\end{center}
\end{figure}

In this subsection, we compare the inelastic mean free path $l$ induced by $e$-$e$ interaction with the elastic mean free path $l_e$ due to charged impurity in the environment. The ratio $\zeta = l/l_e \gg 1$ is the necessary condition for quantum interference effects to be operational in experiments since interference necessarily requires the phase coherence of energy eigenstates. A detailed discussion on this issue has been given in Ref.~[\onlinecite{DasHwangLi_PRB12}], where, however, the inelastic mean free path was not calculated. In Fig.~\ref{fig:compa}, we show our calculated ratio $\zeta$ as a function of temperature for different values of the potential fluctuation (i.e. puddle) parameter $s$. This parameter is defined as the standard deviation of the probability distribution of the disorder potential at a given point in the graphene plane. It can be tuned using screening by the second nearby graphene layer\cite{DasHwangLi_PRB12}. A detailed calculation for the elastic mean free path $l_e$ has been given in Ref.~[\onlinecite{qzli_PRB11}]. Note that we do not take into account electron-phonon scattering in the calculation of the mean free path here since the temperature ($T< 100 K$) is relatively low and the inelastic mean free path can be even shorter if we include electron-phonon scattering mechanism. We defer our discussion of electron-phonon interaction to section \ref{sec:eph} of the paper.

\subsection{Inelastic mean free path and imaginary part of self-energy for intrinsic graphene}

In this subsection, we show our numerical results for the inelastic mean free path $l$ and the imaginary part of the self-energy for intrinsic graphene. The behavior of inelastic $e$-$e$ scattering of intrinsic graphene is quite different from that of extrinsic graphene. Intrinsic graphene is a marginal Fermi liquid\cite{DasHwangTse_PRB07}. By contrast,  extrinsic graphene is a well-defined and relatively weak-coupling Fermi liquid since typically $r_s<1$ and is independent of carrier density.

\begin{figure}
\begin{center}
\includegraphics[width=0.99\columnwidth]{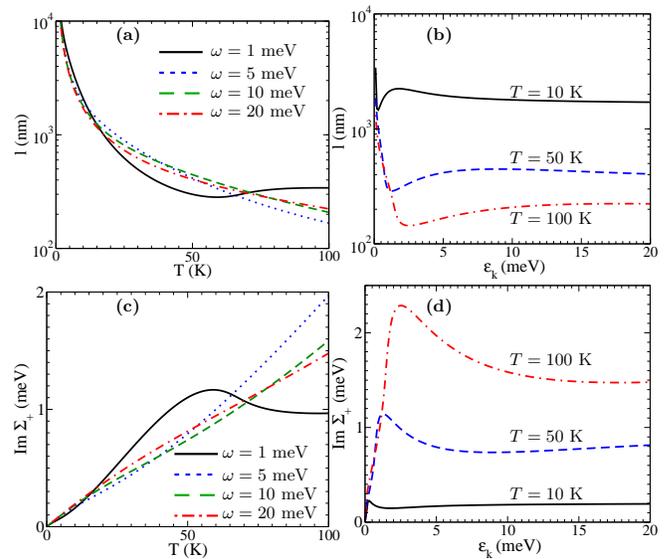}
\caption{ (Color online). Calculated inelastic scattering length $l$ and the associated on-shell imaginary part of the self-energy $\text{Im} \Sigma_{+}^{R}(k, \epsilon_k)$ for intrinsic graphene  ($\varepsilon_F \equiv 0$) with dielectric constant $\kappa = 5$. (a) $l$ as a function of temperature $T$ for different energy $\varepsilon_{\bk}$.  The solid, dotted, dashed and dot-dashed  lines are for energy  $\varepsilon_{\bk} = 1$ meV, $5$ meV, $10$ meV, and $ 20$ meV. (b) $l$ as a function of energy $\omega$ for different temperatures. The solid, dashed and dot-dashed  lines are for temperature $T= 10$ K, $50$ K, and $100$ K. (c) and (d) are the associated $\text{Im} \Sigma_{+}^{R}(k, \epsilon_k)$ corresponding to (a) and (b), respectively.
}
\label{fig:intiml}
\end{center}
\end{figure}

In order to compare with the results provided for extrinsic graphene, we show the finite-$T$ inelastic scattering mean free path $l$ for intrinsic graphene in Fig.~\ref{fig:intiml}(a), (b) and the corresponding imaginary part of the self-energy in Fig.~\ref{fig:intiml}(c), (d) using similar parameters as used for our extrinsic graphene results. Similar to extrinsic graphene, the inelastic mean free path of intrinsic graphene is a monotonically decreasing function of temperature. However, the leading order temperature dependence of $\text{Im}\Sigma^R_+(\bk,\varepsilon_{\bk})$ for intrinsic graphene is linear in $T$ as shown in Fig.~\ref{fig:intiml}. At zero temperature, $\text{Im}\Sigma^R_+(\bk,\xi_{\bk})$ vanishes due to phase-space restrictions leading to infinite  $l$. On the other hand, finite temperature effects give rise to a nonmonotonic, energy dependent $l$ and $\text{Im} \Sigma^R_+(\bk,\xi_{\bk})$. $\text{Im} \Sigma^R_+(\bk,\varepsilon_{\bk})$ (or the quasiparticle scattering rate) first increases linearly with $\varepsilon_{\bk}$ and then decreases with $\varepsilon_{\bk}$. This nonmonotonic energy dependent feature of $\text{Im} \Sigma^R_+(\bk,\xi_{\bk})$ has also been found in Ref.~[\onlinecite{Shutt_PRB11}], and can be explained by the peculiar property of the dynamic dielectric function of intrinsic graphene at finite temperatures.

\begin{figure}
\begin{center}
\includegraphics[width=0.99\columnwidth]{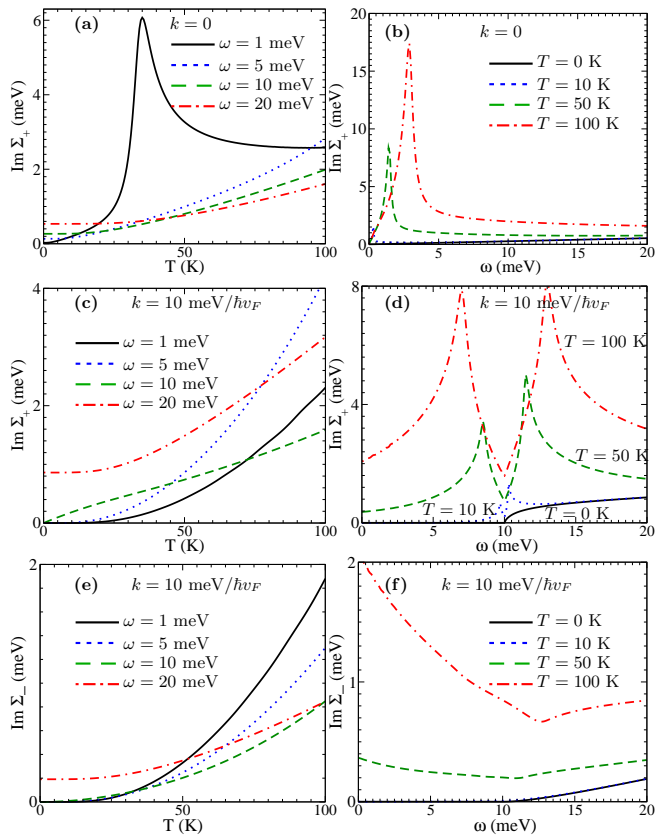}
\caption{(Color online).  Calculated imaginary part of the self-energy Im$\Sigma^{R}(k, \omega)$ for intrinsic graphene (the chemical potential $\mu \equiv 0$) with $\kappa = 5$. (a) and (b), Conduction band Im$\Sigma_{+}^{R}(k=0, \omega)$ as a function of temperature $T$ and energy $\omega$, respectively. (c) and (d), Conduction band Im$\Sigma_{+}^{R}(k, \omega)$ with $k= 10$ meV$/\hbar v_F$ as a function of temperature $T$ and energy $\omega$, respectively. (e) and (f) Valence band Im$\Sigma_{-}^{R}(k, \omega)$ with $k= 10$ meV$/\hbar v_F$ as a function of temperature $T$ and energy $\omega$, respectively. The solid, dotted, dashed and dot-dashed lines in Figs. (a), (c) and (e) are for energy  $\omega = 1$ meV, $5$ meV, $10$ meV, and $ 20$ meV. While the solid, dotted, dashed and dot-dashed lines in Figs. (b), (d) and (f) are for temperature $T = 0$ K, $10$ K, $50$ K, and $100$ K. Note that $\text{Im} \Sigma_{+}^{R}(k=0, \omega) =\text{Im} \Sigma_{-}^{R}(k=0, \omega) $.
}
\label{fig:intk0k10}
\end{center}
\end{figure}

The numerical results for the off-shell $\text{Im} \Sigma^R(k,\omega)$ for intrinsic graphene are shown in Fig.~\ref{fig:intk0k10}. Fig.~\ref{fig:intk0k10}(a) presents $\text{Im}\Sigma^R_{+}(0,\omega)$ as a function of temperature $T$. Quite different from extrinsic graphene, where the plasmon peaks are broadened by Landau damping due to finite temperature effects, intrinsic graphene has no plasmon excitations at zero temperature as seen from Eq.~\eqref{eq:intpola}. Note, however, that the plasmon modes are thermally restored at finite temperatures in intrinsic graphene. We find that there is a peak in $\text{Im}\Sigma^R(0,\omega)$ as a function of temperature, which arises from the appearance of thermal plasmon modes. In particular, the position of these peaks is approximately proportional to the energy $\omega$ because the rescaled function $\text{Im} \tilde{\Sigma}^R(\tilde{k},\tilde{\omega})$ is universal with respect to the rescaling, $\tilde{\Sigma}^R = \Sigma^R/(k_B T)$, $\tilde{k} = \hbar v_F k/(k_B T)$ and $\tilde{\omega} = \omega / (k_B T)$. Only the line of $\text{Im} \Sigma^R(0,\omega)$ with $\omega = 1$ meV in Fig.~\ref{fig:intk0k10}(a) shows a peak since the position of peaks for higher energy lie outside of the temperature regime we are interested in. Fig.~\ref{fig:intk0k10}(b) shows the energy dependence of $\text{Im} \Sigma^R_{+}(0,\omega)$ for several values of temperature. At zero temperature, $\text{Im} \Sigma^R_{+}(0,\omega)$ is linearly increasing as a function of $\omega$ indicated in Eq.~\eqref{eq:intt0}. In the lower energy regime, $\text{Im} \Sigma^R_{+}(0,\omega)$ shows a peak associated with the resonance of plasmon excitations. Specifically, both the position and the height of the peak are roughly proportional to temperature $T$ as mentioned above. The finite temperature $\text{Im} \Sigma^R_{+}(0,\omega)$ approaches the zero temperature $\text{Im}\Sigma^R_{+}(0,\omega)$ as the energy $\omega$ further increases. According to Fig.~\ref{fig:intk0k10}(c), the temperature dependence of off-shell conduction band $\text{Im} \Sigma^R_{+}(k,\omega)$ is exponentially suppressed while the on-shell $\text{Im} \Sigma^R_{+}(k,\omega)$ is proportional to $T$ at low temperatures. We plot the conduction band $\text{Im} \Sigma^R_{+}(k,\omega)$ as a function of energy in Fig.~\ref{fig:intk0k10}(d). One interesting feature is that $\text{Im} \Sigma^R_{+}(k,\omega)$ vanishes for $\omega < \varepsilon_{\bk}$  and it increases rapidly with $\omega$ for $\omega > \varepsilon_{\bk}$ at zero temperature. As the temperature increases, two characteristic peaks show up in $\text{Im} \Sigma^R_{+}(k,\omega)$ similar to extrinsic graphene as shown in Fig.~\ref{fig:exim}(b). This arises from an interplay between plasmon enhancement and thermally smeared single particle continuum boundaries. We have found that the positions of these peaks $|\omega_{pl}- \varepsilon_{\bk}|$ are proportional to the temperature $T$. We provide the valence band $\text{Im} \Sigma^R_{-}(k,\omega)$ as a function of temperature $T$ and energy $\omega$ in Figs.~\ref{fig:intk0k10}(e) and (f), respectively. The low temperature correction to  $\text{Im}\Sigma^R_{-}(k,\omega)$ is exponentially suppressed for $\omega > 0$ due to the constraint imposed by the Fermi and Bose distribution function in Eq.~\eqref{eq:exim}. $\text{Im} \Sigma^R_{-}(k,\omega)$ increases rapidly at higher temperatures, especially for small energy $\omega$. From Fig.~\ref{fig:intk0k10}(f), we can see that $\text{Im} \Sigma^R_{-}(k,\omega)$ increases slowly with $\omega$ for $\omega > \varepsilon_{\bk}$ at low temperatures and $\text{Im} \Sigma^R_{-}(k,\omega)$ shows non-monotonic dependence on $\omega$ at higher temperature. This non-monotonicity can be understood through the competition between a decrease in valence band contribution ($s'=-1$) and an increase in conduction band contribution ($s'=1$) with increasing energy $\omega$.

\section{Mean free path due to electron-phonon interaction}
\label{sec:eph}
In this section, we present results for the mean free path $l_{ep}$ induced by electron-phonon interaction. We analytically calculate its high-($T \gg T_{BG}$) and low-temperature ($T \ll T_{BG}$) limits, while in the intermediate temperature regime we provide numerical results. In addition, we consider the energy and carrier density dependence of $l_{ep}$ for different values of temperature. In this work, we consider acoustic phonons only, and  neglect the optical phonons because their energy in graphene is too high ($\sim 2000$ K) for them to play any important role in the temperature range ($T \lesssim 500$ K) of our interest.

\subsection{Imaginary part of self-energy due to electron-phonon interaction}

The  mean free path $l_{ep}$ is directly related to the on-shell imaginary part of the self-energy, i.e. $l_{ep} = v_F / (2 \text{Im} \Sigma^{ep}_{+} (k,\xi_{k}))$, as given in Sec.~\ref{sec:theory}.  Thus, we first present the theoretical formalism for calculating $\text{Im} \Sigma^{ep}_{s} (k,\omega)$. We consider the electron-phonon interaction through the deformation potential coupling. In this case,  $\text{Im} \Sigma^{ep}_{s} (k,\omega)$ is given by\cite{Mahan}:
\begin{eqnarray}
&&\text{Im} \Sigma^{ep}_{s} (k,\omega) = \pi \sum_{s', \nu} \int \frac{d^2 q}{(2 \pi)^2} |M|^2 \frac{1+ s s' \cos\theta}{2} \nonumber
\\
&&\times  \left[n_F(\omega_q + \nu \omega) + n_B(\omega_q)\right]\delta(\omega + \nu \omega_q - \epsilon'),
\label{eq:imeph}
\end{eqnarray}
where $|M|^2 = \frac{D^2 \hbar q}{ 2 \rho_m v_l}$ is the electron-phonon scattering matrix element, while $D$, $\rho_m$ and $v_l$ are the deformation potential, the graphene mass density and the phonon velocity, respectively. $s, s' = \pm 1$ denote the band indices, and $\nu = 1$ ($-1$) corresponds to the absorption (emission) of an acoustic phonon with  frequency $\omega_q = v_l q$. $\epsilon' = s' \hbar v_F |\bk + \bq| - \mu$, and $\theta$ is the angle between $\bk$ and $\bk+\bq$. 

\subsection{The asymptotic behavior of $\text{Im} \Sigma^{ep}_{s} (k,\xi_k)$}
We start with the asymptotic behavior  for the energy- and temperature-dependence of $\text{Im} \Sigma^{ep}_{s}$, calculated analytically within the quasielastic scattering approximation\cite{HwangDasPhonon_PRB08,MinHwangDas_arX10,Kristen_PRB12},  namely $|\bk| \sim |\bk+\bq|$. This approximation is justified by the fact that the phonon velocity is much smaller than the Fermi velocity $v_F$ in graphene.

At zero temperature, $n_F$ becomes the Heaviside unit step function, $n_B(\omega_q) \equiv 0$, and the interband scattering vanishes. Then, the integration in Eq.~\eqref{eq:imeph} can be carried out in the limit $\xi_k \ll E_F$ as:
\beq
\text{Im} \Sigma^{ep}_{+} (k,\xi_{k}) \simeq \frac{1}{8 \pi} \frac{D^2 k_F^2}{\rho_m v_l v_F} \left(\frac{v_F}{v_l}\right)^{2}\left(\frac{\xi_k}{E_F}\right)^{2}
\label{eq:phanaimt0}
\eeq
which shows that $\text{Im} \Sigma^{ep}_{+} (k,\xi_{k})$ increases quadratically with increasing quasiparticle energy, in contrast to $\text{Im} \Sigma_{+} (k,\xi_{k}) \propto \xi_{k}^2 \ln \xi_{k}$ induced by $e$-$e$ interaction. Eq.~\eqref{eq:phanaimt0} implies that $l_{ep} \propto \xi_k^{-2}$ at $T=0$.

Next, we consider the temperature dependence of $\text{Im} \Sigma^{ep}_{+} (k,\xi_{k})$. In the low temperature Bloch-Gr\"{u}neisen limit, $T \ll T_{BG}$, with $T_{BG} = 2 k_F v_l/k_B$, where the phonon system is degenerate, we get the following asymptotic form for the on-shell $\text{Im} \Sigma^{ep}_{+} (k, \omega)$:
\beq
\text{Im} \Sigma^{ep}_{+} (k_F,\xi_{k_F}) \simeq \frac{\pi}{2} \frac{D^2 k_F^2}{\rho_m v_l v_F}\left(\frac{T}{T_{BG}}\right)^{2}
\label{eq:phanaimlow}
\eeq
In the Bloch-Gr\"{u}neisen regime, the temperature dependence of $l_{ep}$ is a power law with $l_{ep} \propto T^{-2}$. We also find that $l_{ep}$ does not depend on the  carrier density in the low temperature limit.

On the other hand, in the high temperature regime $T_F \gg T \gg T_{BG}$, where the phonon follows the nondegenerate equipartition distribution, the asymptotic formula for the on-shell $\text{Im} \Sigma^{ep}_{+} (k,\omega)$ is given by:
\beq
\text{Im} \Sigma^{ep}_{+} (k_F,\xi_{k_F}) \simeq \frac{1}{2} \frac{D^2 k_F^2}{\rho_m v_l v_F} \frac{T}{T_{BG}}
\label{eq:phanaim}
\eeq
Eq.~\eqref{eq:phanaim} shows that $l_{ep}$ decreases inverse linearly with increasing temperature, and  is proportional to $n^{-1/2}$. We note that $T_{BG} \sim k_F \sim \sqrt{n}$ has a carrier density dependence and increases with increased doping in the system.

\subsection{Numerical results of $l_{ep}$}
\begin{figure}
\begin{center}
\includegraphics[width=0.99\columnwidth]{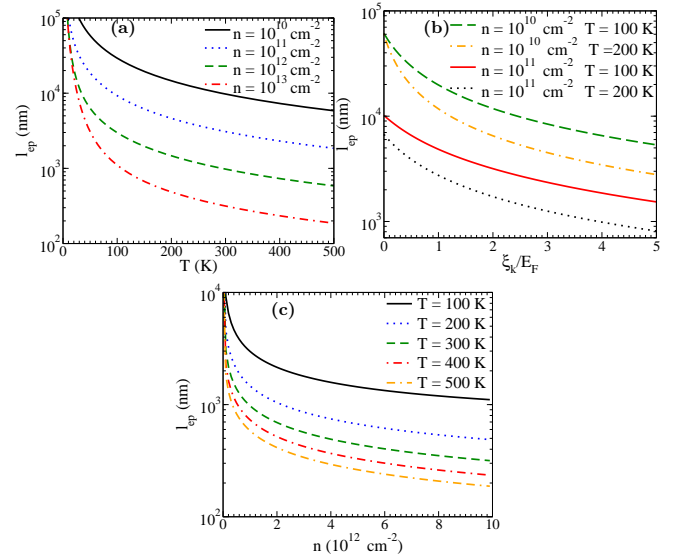}
\caption{(Color online).  Calculated mean free path $l_{ep}$ induced by electron-phonon interaction. We use \cite{MinHwangDas_arX10} $D= 25$ eV, $v_l = 2.6 \times 10^{6}$ cm/s and $\rho_m = 7.6 \times 10^{-8}$ g/cm$^{2}$. (a) $l_{ep}$ for $\xi_k=\xi_{k_F}$  as a function of temperature $T$ for different carrier densities. (b) $l_{ep}$ as a function of energy $\xi_k/E_F$ for different carrier densities and temperatures. (c) $l_{ep}$ for $\xi_k = \xi_{k_F}$ as a function of carrier density for different temperatures.  }
\label{fig:eph}
\end{center}
\end{figure}

Fig.~\ref{fig:eph} shows our numerical results for the mean free path $l_{ep}$. Fig.~\ref{fig:eph}(a) plots $l_{ep}$  as a function of temperature for different values of carrier density. It is clear that $l_{ep}$ decreases monotonically with increasing temperature. Fig.~\ref{fig:eph}(b) and (c) demonstrate that $l_{ep}$ is a decreasing function of $\xi_k$ and the carrier density $n$. From Fig.~\ref{fig:eph}(c), we see that $l_{ep}$ is longer than $1000$ nm for $T<100$ K and $n < 10^{13}$ cm$^{-2}$. Comparing Fig.~\ref{fig:eximl} (a) with Fig.~\ref{fig:eph} (a), we find that $l_{ep}$ is about one order of magnitude longer than the inelastic mean free path $l$  induced by $e$-$e$ interaction for $T <100$ K and $n \lesssim 10^{12}$ cm$^{-2}$. Thus, the inelastic scattering is dominated by  electron-electron scattering processes in this regime. With increasing carrier density and temperature, $l_{ep}$ becomes comparable to $l$, implying a considerable contribution of $e$-$p$ interaction to the inelastic scattering processes. We mention that in the presence of both $e$-$e$ and $e$-$p$ interaction, the net quasiparticle scattering rate is given by $1/\tau_{ee} + 1/\tau_{ep}$ and thus the net inelastic mean free path goes as $\sim 1/(1/l_{ee} + 1/l_{ep})$.

\section{Discussion and Conclusion}
\label{sec:conclu}

Before concluding, we first discuss the similarity and difference between graphene and conventional 2DES. Because of the gapless nature of graphene, there are two contributing bands in graphene, i.e., conduction band and valence band, which allow both interband and intraband single particle excitations. By contrast, the conventional 2DES has only one contributing band, either the conduction or the valence band, because of the large energy gap. In addition, at zero temperature,  $\text{Im}\Sigma^R_+ (\bk,\xi_{\bk})$ of graphene is a smooth function of $\xi_{\bk}$, where both plasmon and interband excitations are absent. While there is a discontinuity in $\text{Im}\Sigma^R (\bk,\xi_{\bk})$ for a 2DES, caused by the plasmon\cite{Giuliani_PRB82} and electron-hole excitations\cite{Jalabert_PRB89,HwangHuInelastic_PRB07}, there is no such discontinuity in graphene due to its gaplessness. 

We now discuss the difference between extrinsic and intrinsic graphene. The low-energy  quasiparticle scattering rate of extrinsic graphene is quite similar to conventional 2DES, which is well described by the Fermi liquid theory. The non-monotonic temperature dependence of $\text{Im} \Sigma^R_+ (\bk,\omega)$ in extrinsic graphene arises from the competition between plasmon broadening effects and thermally smeared single particle excitation boundaries. On the other hand, the quasiparticle scattering rate of intrinsic graphene is linear in energy $\omega$,  a signature of the marginal Fermi liquid behavior. The characteristic feature of the non-monotonic $\text{Im} \Sigma^R_+ (\bk, \omega)$ for intrinsic graphene as a function of temperature is due to an interplay between plasmon enhancement and enlarged single particle excitation continua. We have also compared the inelastic mean free path with the elastic mean free path in order to assess the density and temperature range where the inelastic scattering length is decisively longer so that quantum interference is in principle allowed in graphene. When $l < l_e$, quantum interference induced localization effects cannot manifest itself.  In addition, motivated by a recent experiment on double-layer graphene, we have obtained analytic results for inelastic scattering in double-layer graphene where one layer acts to screen the other layer.

Our numerical calculation shows that inelastic scattering due to electron-phonon interaction is negligible for $T < 100$ K, and $e$-$e$ interaction dominates in this temperature range. The electron-phonon scattering starts to play a role in extrinsic graphene when the temperature is above $200$ K. In particular, the temperature dependence of $l_{ep}$ changes from $T^{-2}$ to $T^{-1}$ as the temperature increases from the low-$T$ Bloch-Gr\"{u}neisen regime to the high-$T$ equipartition regime. On the other hand, the low-temperature asymptotic temperature dependence of the electron-electron interaction leads to $l \sim (T^2 \ln T)^{-1}$ for extrinsic graphene and $l \sim T^{-1}$ for intrinsic graphene.

We should note here that our calculation of the inelastic scattering length based on the imaginary part of the graphene self-energy is carried out in the so-called ``ballistic limit'' where disorder effects are neglected in the calculation of the inelastic mean free path itself. As such our calculated inelastic mean free path is an upper limit on the phase breaking length $l_\phi$ applicable to graphene quantum interference phenomena. It is well-known that disorder has qualitative and quantitative effects on the inelastic mean free path\cite{BelitzDas_PRB87} and the phase breaking length, in general suppressing the mean free path substantially from its ballistic limit. In the diffusive limit, we anticipate the inelastic mean free path due to $e$-$e$ interaction to have the asymptotic low-temperature behavior of $l \sim (T \ln T)^{-1}$ instead of the $l \sim (T^2 \ln T)^{-1}$ ballistic behavior we find in Sec.~\ref{sec:theory}. The phase breaking length $l_\phi$ is anticipated to be $l_\phi \sim T^{-1}$ in the low-temperature diffusive regime. Thus, the actual inelastic length due to electron-electron interaction would be smaller than the ballistic limit results we obtain in the current work. We note that the disorder effect on the electron-phonon interaction induced inelastic mean free path $l_{ep}$ derived in Sec.~\ref{sec:eph} is likely to be small\cite{BelitzDas_PRB87}, and our ballistic limit results should apply equally well to the diffusive regime also.

To conclude, we have obtained analytical asymptotic behavior of $\text{Im}\Sigma^R_+(\bk,\xi_{\bk})$ in graphene utilizing the GW approximation. We have also analyzed $\text{Im}\Sigma^R_+(\bk,\xi_{\bk})$ for a double-layer graphene system. We have shown numerical results for the inelastic mean free path $l$ and $\text{Im} \Sigma^R (\bk,\omega)$. We emphasize that finite temperature has strong effects on the inelastic scattering mean free path. We have discussed the difference between graphene and conventional 2DES. We have also provided results for the inelastic mean free path arising from electron-phonon interaction which becomes important only at very high temperatures since the typical electron-phonon dimensionless coupling constant ($<0.04 \ \text{for carrier density} \ n< 10^{13} \ \text{cm}^{-2})$ is much less than the typical dimensionless coupling constant $r_s (\sim 0.4)$ in graphene.  Our use of the GW approximation, which is the leading order Feynman-Dyson perturbative expansion in the dynamically screened Coulomb interaction, should be an excellent approximation for graphene (with quantitative predictive power) because of the small values of $r_s (<1)$ in graphere on substrates.  Going beyond the GW approximation is a formidable task which may not be necessary in understanding graphene inelastic processes.

\begin{acknowledgments}

QL acknowledges helpful discussions with Euyheon Hwang.  This work is supported by US-ONR-MURI and NRI-SWAN.

\end{acknowledgments}


%

\end{document}